# ABNORMAL TEMPERATURE-DIFFUSION RELATIONSHIP IN THE EXTERNAL PEREODIC FIELDS

## I.G. Marchenko[1,2] and I.I. Marchenko[3]

[1]*National Scientific Center "Kharkov Institute of Physics and Technology"*
*1, Akademicheskaya St., Kharkov, 61108, Ukraine*
*E-mail: march@kipt.kharkov.ua*
[2]*Kharkov National University, 4 Svobody Sq., Kharkov, 61077, Ukraine*
[3]*NTU „Kharkov Polytechnic Institute", 21 Frunze St., Kharkov, 61145, Ukraine*
(Dated: October 4, 2011)

Using the methods of computer modeling this scientific paper studies the special features of diffusion of the particles subjected to the external periodic force in the crystal lattice. The particle motion is described by a Langevin equation. The systems with a low friction coefficient may experience abnormal diffusion modes, in particular hyperdiffusion and subdiffusion. The applied external time-periodic field causes limitation of time intervals of abnormal diffusion making the diffusion coefficients dependent on frequency of applied force. The temperature relationships of these values have been calculated. It has been shown that the diffusion coefficients behave in abnormal ways as the temperature changes. In some temperature intervals the diffusion may increase as the temperature drops. Location and width of these intervals depend on the frequency of the external field.

Transport and diffusion of particles have been subjects of in-depth studies for many decades. In his basic studies [1] Einstein established the relationship for the diffusion coefficient $D$ of the Brownian particles moving in the viscous medium subjected to the action of external force:

$$D = kT\mu, \qquad (1)$$

where $T$ is a temperature, $k$ is a Boltzmann constant and $\mu$ is a particle mobility. The diffusion of atoms and defects in the crystals are traditionally described by the Arrhenius dependence [2], which is in good conformance with the obtained experimental data. The elementary event of diffusion is viewed as the negotiation of activation barrier by a particle. The probability of such an event is usually determined using Kramer's rule [3], which binds diffusion with a probability of overcoming the activation barrier $U$:

$$D = D_0 \exp(-U/kT). \qquad (2)$$

The expression obtained for the diffusion coefficients is based on many simplifying theoretical assumptions. The complicated dynamics of a particle being taken into account in a potential pattern can result in some other temperature relationships. It is shown in paper [4] that the diffusion for the low friction coefficient $\gamma$ systems is described by the relationship

$$D \sim \frac{T}{\gamma} \exp(-U/kT). \qquad (3)$$

Other options of temperature dependence are also possible. However, if the physical mechanisms of diffusion are not changing it always becomes stronger with an increase in temperature. It corresponds to our intuitive ideas of that the material "mixing" rate increases as the temperature rises.

We will show in this paper that the influence produced by the time-periodic external field onto a system may reduce the diffusion of particles in the crystal structures with an increment in temperature.

The migration of particles in the spatial periodic potentials subjected to the action of external force has vigorously been studied over the last few years [5-8]. The previous research showed that the systems with low $\gamma$-values may develop a diffusion that differs from the ordinary one and it is called subdiffusion or hyperdiffusion [7, 9]. In this case the attention of the researchers was focused on the conditions of emergence of the ordered motion of particles subjected to the action of external force. However, the temperature dependence of mobility requires further studies because it is of great interest from the practical standpoint.

Our purpose has been to study the temperature peculiarities of particle diffusion in a crystal lattice subjected to the action of external periodic field, using the methods of computer modeling.

In order to demonstrate the main physical results we studied the case of one–dimensional diffusion. General conclusion drawn for the case of the surface and volumetric diffusion does not change the main physical findings of this scientific paper.

The motion of particle inside the one-dimensional lattice was described by a Langevin equation [10]:

$$m\ddot{x} = -\frac{\partial}{\partial x}U(x) - \gamma \dot{x} + F_t(t) + \xi(t), \qquad (4)$$

where $t$ is a time, $x$ is a particle coordinate in the one-dimensional lattice, $m$ is a mass and $\gamma$ is a friction coefficient. The overdots denote a derivative with respect to $t$. The thermal fluctuations $\xi(t)$ represent uncorrelated white noises that obey the relationship

$$\langle \xi(t)\xi(t') \rangle = 2m\gamma kT\delta(t-t'). \qquad (5)$$

The potential energy of a particle in the crystal lattice field $U$ was described by the following expression

$$U = -\frac{U_0}{2}\cos\left(\frac{2\pi}{a}x\right), \qquad (6)$$

where $a$ is a lattice constant and $U_0$ is a height of potential barrier.

A moving particle was subjected to the action of the force $F$ produced by a lattice:

$$F = -\frac{\partial U}{\partial x} = F_{cr}\sin\left(\frac{2\pi}{a}r\right); \qquad (7)$$

hereinafter the value of $F_{cr} = \frac{\pi}{a}U_0$ will be referred to as the critical force [7]. The time-periodic external field was described by the following expression:

$$F_t(t) = F_0 \sin(\omega t), \qquad (8)$$

where $\omega$ is an angular frequency of external force and $F_0$ is an amplitude of it.

A hydrogen atom was selected to be a migrating particle. The value of activation barrier was 8meV, which is a typical value for the diffusion of adatoms across the close-packed surfaces of metals with the face-centered cubic structure (fcc) and hexagonal close-packed (hcp) structure as well as for the diffusion of hydrogen within the material volume [10]. The lattice constant $a$ was selected to be equal to 2 Å.

The diffusion coefficient was calculated using the traditional approach:

$$D = \lim_{t\to\infty} D_\tau(t) = \lim_{t\to\infty} \frac{\langle (x(t) - \langle x(t) \rangle)^2 \rangle}{2t}, \qquad (9)$$

where the angle brackets $\langle ... \rangle$ denote the ensemble averaging.

The motion equations were numerically solved with a time increment of $\Delta t \approx 1$ fs. The statistical ensemble averaging was done for the amount of particles $N = 4 \cdot 10^4$. The initial conditions were specified in the following way: a particle was placed at the origin of coordinates and it was randomly given a velocity having the Maxwell temperature distribution. The system was then subjected to thermalization during $10^4$ time steps. Afterwards the particle moved at acquired velocity to the first unit cell.

While analyzing the results of modeling it is reasonable to write the equation in terms of scaled dimensionless variables, in particular, time $\tau$, distance $x'$ and temperature $T'$ [4]:

$$\tau = t\frac{1}{a}\sqrt{\frac{U_0}{m}}\ ;\quad x' = \frac{x}{a}\ ;\quad T' = \frac{kT}{U_0}\ . \tag{10}$$

The dimensionless friction coefficient will be equal to

$$\gamma' = \frac{\gamma a}{(mU_0)^{1/2}}\ , \tag{11}$$

and was equal to 0.2 in present calculations. This value corresponds to the value observed in the experiment carried out to study the hydrogen diffusion on the platinum surface [11]. Under action of constant force ($\omega=0$) different modes of particle motion (Fig.1) are realized depending on $F$ and $F_{cr}$ relationship. The obtained results are well-correlated with the data obtained by the other authors [9]. The plot gives the dimensionless values of mean-square displacements $\sigma^2 = \langle (x'(t) - \langle x'(t) \rangle)^2 \rangle$ in the logarithmic scale.

The first plot corresponds to the case when the external force is not applied ($F=0$). The particle motion occurs only under action of thermal fluctuations. It is seen from the Figure that the mean-square displacements as a function of time $\sigma^2 \sim t$ have in this case a linear relationship, which corresponds to the common diffusion [3]. An increase in force (curve 2) leads to the appearance of short area of ballistic diffusion, which is characterized by the relationship of $\sigma^2 \sim t^2$. This transient mode of diffusion is only observed at short times ($t/\Delta t < 5\cdot 10^3$). With a further increase in force (curve 3) the area of hyperdiffusion appears which is characterized by cubic dependence of $\sigma^2$ on time.

However, at extended times of ($t/\Delta t > 10^6$) the dispersion continues to display a linear dependence on time. The successive increase in the external force (curve 4) results in the appearance of a new type of particle motion, in which $\sigma^2$ is not dependent on time. It is seen from the plot that this mode appears after the hyperdiffusion phase. With a further increase in $F$ the area of diffusionless transport expands (curve 5). With an increase in the external force the interval of hyperdiffusion reduces and the interval of diffusionless transport decreases. When the force reaches the value that exceeds the critical one (curve 6) the transient states are actually smoothed out and diffusion is again described by the linear dependence of $\sigma^2$ on time, as in the case of absence of external force. It is seen from the plots that the curve 6 is parallel to the curve 1 at $t/\Delta t > 5\cdot 10^3$; however, the curve 6 is offset relative to it by two orders of magnitude in the Y-axis. A further increase in F results in the recovery of the mode of common diffusion in the entire time interval under consideration.

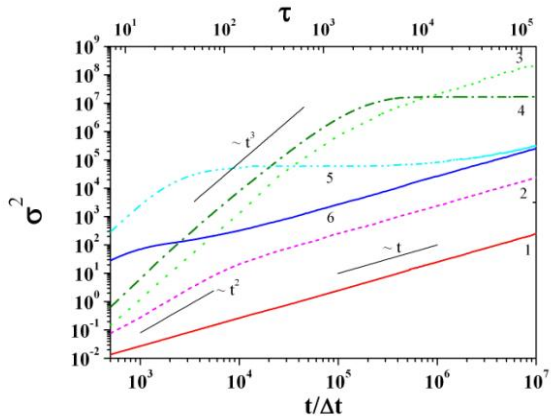
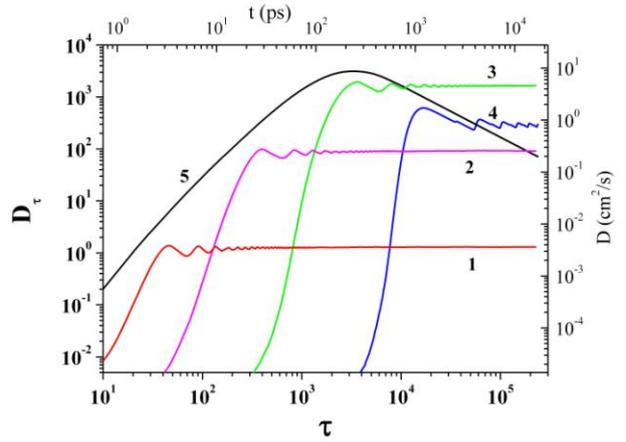

**Fig. 1.** Mean-square particle displacement as a function of time for different values of permanent force $F$. Temperature T=120K ($T' \approx 0.13$). Curve 1 - $F'=0$, 2 - $F'=0.05$, 3 - $F'=0.09$, 4 - $F'=0.15$, 5 - $F'=0.7$, 6 - $F'=1.0$. The lengths of solid line show the dependences $\sim t$, $\sim t^2$ and $\sim t^3$.

**Fig. 2.** Diffusion coefficient as a function of time for different frequencies $\omega'$ at a temperature of $T=120K$ ($T'=0.13$). The periodic force amplitude is $F_0 = 0.15 F_{cr}$ 1 - $\omega' = 7.1\cdot 10^{-2}$ ($\nu = 156$ GHz); 2 - $\omega' = 7.1\cdot 10^{-3}$ ($\nu = 15.6$ GHz); 3 - $\omega' = 7.1\cdot 10^{-4}$ ($\nu = 1.56$ GHz); 4 - $\omega' = 7.1\cdot 10^{-5}$ ($\nu = 156$ MHz); 5 - $\omega' = 0$ (constant force).

The applied external periodic field leads to the time limitation of diffusion processes. As a result the modes of abnormal diffusion are realized that are peculiar for the time equal to the period of external field oscillations. Fig. 2 gives the plots for the time change of coefficients $D_\tau(t)$ at different frequencies of applied external field. The Figure legends give both the values of angular frequencies $\omega'$ in dimensionless units and the frequency of the external field in Hz. The Figure shows that the coefficients $D_\tau(t)$ reach the saturation level at extended times; it is indicative of traditional diffusion, which is characterized by the linear relationship of the root-mean –square time offsets. At the same time the value of diffusion coefficient depends considerably on the frequency of the external field.

A coarse estimate shows that the steady-state diffusion coefficient for the given frequency is approximately equal to the value of $D_\tau(t)$, which is observed under action of constant force at the time moment equal to one oscillation period of external field. In this connection the diffusion coefficient $D_\nu$ first increases with an increase in frequency and then decreases following the curve 1. Such a nonlinear behavior of a diffusion coefficient may change with a change in temperature.

Fig. 3 gives the plots of relationships $D_\tau(t)$ for the different temperatures at a fixed frequency of external field $\omega' = 7.1 \cdot 10^{-3}$. It is seen from the Figure that the steady-state diffusion coefficient $D$ increases in the temperature range of $T' = 0.07$ to $T' = 0.26$. With a further increase in temperature the value of $D$ decreases. Such an unusual behavior of particle diffusion in the periodic field shows that the temperature dependences of diffusion may show the abnormal zones, in which the diffusion coefficient may increase with a temperature drop.

Fig. 4 gives the temperature relationships of diffusion coefficients $D$ for different frequencies of external field. The round markers (curve 1) give the values of diffusion coefficients when the external force is not applied. The dashed line shows the Arrhenius relationship. It follows from the plot that the diffusion is described by the common exponential dependence on time when the external force is not applied. The markers (curve 2) give the values of diffusion coefficient when the particle is subjected to the action of constant force $F_0 = 0.15 F_{cr} (\nu = 0)$. In this case the temperature behavior of the diffusion is well-approximated by the relationship (1) for the motion of Brownian particle in the viscous media, which is shown in the plot by a dashed line.

The plots 3 to 6 show that the temperature relationships of diffusion coefficients experience significant changes with a decrease in frequency. In the domain of high frequencies (curve 3) the diffusion increases with an increase in temperature. Such traditional behavior is also observed at very low frequencies (curve 2). In the intermediate frequency range of external field the diffusion coefficient has a no uniform dependence on temperature. The tenfold increase in frequency (see curve 4) results in the appearance of maximum in the temperature relationship of diffusion coefficient at $T' \approx 0.2$. Such an abnormal behavior of diffusion coefficient, which is characterized by more intensive diffusion at temperature drop, becomes more potent as the frequency decreases. The further decrease in frequency results in the increase of maximum value of diffusion coefficient by one order of magnitude. It shifts to the range of lower temperatures. As the frequency decreases down to $\omega' = 7.1 \cdot 10^{-4}$ the maximum value increases again by more than one order of magnitude. It continues to shift towards the lower temperature values. At such a frequency the maximum value of diffusion coefficient exceeds the coefficient value of thermally activated diffusion by more than nine orders of magnitude in the absence of external forces. Such a considerable increase in diffusion under action of external periodic force can be of practical importance. It is also seen from the Figure that the increase in diffusion takes place in a rather narrow temperature range.

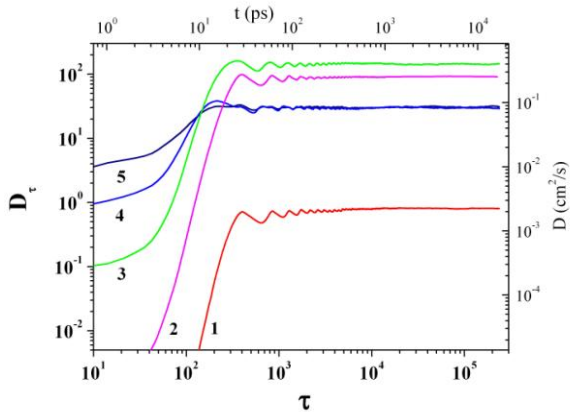 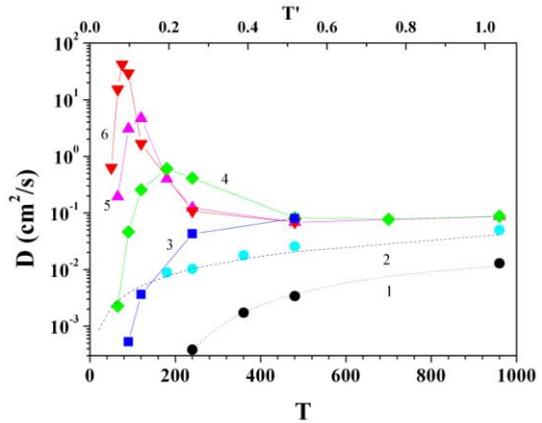

**Fig. 3.** Coefficient $D(t)$ as a function of time at different temperatures. The periodic force amplitude $F_0 = 0.15 F_{cr}$, frequency $\omega' = 7.1 \cdot 10^{-3}$ ($\nu = 15.6$ GHz). 1 – T=65K ($T' = 0.07$), 2 - T=120K ($T' = 0.13$), 3 - T=240K ($T' = 0.26$), 4 - T=480K ($T' = 0.53$), 5 - T=960K ($T' = 1.04$).

**Fig. 4.** Diffusion coefficient $D$ as a function of temperature for different frequencies $\nu$ of external field. Periodic force amplitude is $F_0 = 0.15 F_{cr}$. 2 - $\omega' = 0$; 3 - $\omega' = 7.1 \cdot 10^{-2}$ ($\nu = 156$ GHz); 4 - $\omega' = 7.1 \cdot 10^{-3}$ ($\nu = 15.6$ GHz); 5 - $\omega' = 7.1 \cdot 10^{-4}$ ($\nu = 1.56$ GHz); 6 - $\omega' = 7.1 \cdot 10^{-5}$ ($\nu = 156$ MHz); The relationship of thermal diffusion 1 - $F_0 = 0$, $\nu = 0$. is given for comparison.

The correctness of the above calculations is naturally limited by a lowering of an absolute temperature $T$ below the Einstein $T_E$ (or Debye $T_D$) temperatures for the studied material. If the temperature drops below this value the calculations should be done taking into account the quantum statistics.

This research opens up possibilities for changing the mobility of atoms, point defects and linear defects in the crystals of definite types without changing the solid body temperature. It is very important for the development of new methods used for changing the properties of materials sensitive to thermal action. The fields of application that are of practical importance are as follows: stimulating the enhanced hydrogen release from the fuel elements used for the hydrogen power engineering, annealing the radiation defects in the irradiated crystals, etc.

Integrally this research paper considers the peculiarities of diffusion processes that occur in crystals under the action of the external periodic field. It has been shown that this results in abnormal temperature dependence of particle diffusion. The diffusion can be intensified with a temperature drop. Such an unusual behavior has a nonlinear dependence on frequency of external field. This phenomenon can be experimentally verified and opens up possibilities for changing the mobility of atoms and defects in the crystals of definite types without changing the solid body temperature.